\documentclass[conference,letterpaper]{IEEEtran}
\IEEEoverridecommandlockouts
\usepackage{cite}
\usepackage{amsmath,amssymb,amsfonts}
\usepackage{amsthm}
\usepackage{graphicx}
\usepackage{textcomp}
\usepackage{xcolor}
\definecolor{mygray}{rgb}{0.5,0.5,0.5}
\definecolor{changed}{rgb}{0,0,0}
\definecolor{changed2}{rgb}{0.0,0,0}
\definecolor{myblue}{rgb}{0,0,.6}
\usepackage{textpos}
\usepackage{epstopdf}
\usepackage{comment}
\usepackage{balance}
\usepackage{bm}
\def\BibTeX{{\rm B\kern-.05em{\sc i\kern-.025em b}\kern-.08em
    T\kern-.1667em\lower.7ex\hbox{E}\kern-.125emX}}

\usepackage{subcaption}  %for subfigure environment

\usepackage{hyperref}
\hypersetup{
    pdftitle={HiL Demonstration of Online Battery Capacity and Impedance Estimation with Minimal a Priori Parametrization Effort},    % title
    pdfauthor={Camiel J.J. Beckers, Feye S.J. Hoekstra, Frank Willems},     % author
    pdfkeywords= {State-of-Health (SoH)}{Parameter Estimation}{Battery Management System (BMS)}{Hardware-in-the-Loop (HiL)}{Forgetting-Factor Recursive Least-Squares (FFRLS)}
}

\usepackage{eso-pic}
 \AddToShipoutPictureBG{%
   \AtTextUpperLeft{%
     \put(\LenToUnit{0cm},\LenToUnit{.5cm}){%
         \begin{minipage}[t]{24cm}
           \small
             Accepted author version of {\color{myblue}\href{https://doi.org/10.1109/VPPC63154.2024.10755448}{10.1109/VPPC63154.2024.10755448}}.
           \end{minipage}
     }%
   }%
 }

\begin{document}
\setlength{\baselineskip}{11.8pt}%original = 12
\IEEEpubid{\begin{minipage}{\textwidth}\ \\[50pt] \centering
  \copyright 2024 IEEE.  Personal use of this material is permitted.  Permission from IEEE must be obtained for all other uses, in any current or future media, including reprinting/republishing this material for advertising or promotional purposes, creating new collective works, for resale or redistribution to servers or lists, or reuse of any copyrighted component of this work in other works.
\end{minipage}}

\title{HiL Demonstration of Online Battery Capacity and Impedance Estimation with Minimal a Priori Parametrization Effort
%Note: Sub-titles are not captured in Xplore and should not be used}
\thanks{This work has received financial support from the Dutch Ministry of Economic Affairs and Climate, under the grant ‘R\&D Mobility Sectors’ and as part TNO’s Early Research Program in the scope of the project `AUTO ADAPT'.}
}

%%ALTERNATIVE AUTHOR INDICATION:
\author{\IEEEauthorblockN{Camiel J.J. Beckers\IEEEauthorrefmark{1}, Feye S.J. Hoekstra\IEEEauthorrefmark{1}, Frank Willems\IEEEauthorrefmark{1}\IEEEauthorrefmark{2}
\vspace{0.3cm}
}
\IEEEauthorblockA{\IEEEauthorrefmark{1}Powertrains Dept., TNO Mobility \& Built Environment, Helmond, The Netherlands, Email: camiel.beckers@tno.nl}
\IEEEauthorblockA{\IEEEauthorrefmark{2}
Department of Mechanical Engineering, Eindhoven University of Technology, Eindhoven, The Netherlands}
}

\maketitle

%an abstract of 50-100 words should be written in English
\begin{abstract}
Uncertainty in the aging of batteries in battery electric vehicles impacts both the daily driving range as well as the expected economic lifetime.
This paper presents a method to determine online the capacity and internal resistance of a battery cell based on real-world data.
The method, based on a Joint Extended Kalman Filter combined with Recursive Least Squares, is computationally efficient and does not a priori require a fully characterized cell model.
Offline simulation of the algorithm on data from differently aged cells shows convergence of the algorithm and indicates that capacity and resistance follow the expected trends.
Furthermore, the algorithm is tested online on a Hardware-in-the-Loop setup to demonstrate real-time parameter updates in a realistic driving scenario.
\end{abstract}

\section{Introduction}
%Context
In an effort to reduce greenhouse gas emissions, the world is transitioning to electric mobility.
Many of these new energy vehicles rely on lithium-ion batteries as their primary energy carrier. To ensure safe and reliable operation, a Battery Management System (BMS) monitors the battery cells and enforces strict operational boundaries. One of the challenging aspects of battery management is that the battery capacity and impedance can change and degrade over time and use, which shortens the driving range of the vehicle and can potentially impact safety. Moreover, upcoming EU regulations concerning battery traceability, i.e., the battery passport, require manufacturers to provide up-to-date battery health information. Therefore, accurate tracking of the capacity and impedance is crucial.

Given its importance, the separate topics of impedance and capacity estimation, but in particular also joint, i.e., integrated, solutions have received significant attention in literature. A high variety of solutions exists; Some apply electrochemical impedance-spectroscopy, see, e.g., \cite{su2022fast}, where capacity is linked to particular impedance responses, thus providing a fast capacity estimate. Others apply scanning for particular changes in operational characteristics such as repeated charge curves~\cite{lyu2021partial}. In recent years, many machine learning approaches have also surfaced ranging from straight-forward support vector machines to artificial neural networks~\cite{Sui2021}. Nevertheless, the most described approach for impedance and capacity estimation is some form of a joint state and parameter estimator in combination with a Least Squares (LS) estimator for the capacity based on the difference in estimated State-of-Charge (SoC) and integrated current.

While seemingly straightforward, there exist many combinations or stand-alone implementations of these two options. In this paragraph, a brief impression is provided of the various options. For instance, in \cite{Wu2023}, capacity is estimated using a total LS structure, which shows accurate capacity estimation, but presents an unfair comparison with ordinary LS and requires a significant number of evaluations. Similarly, capacity is estimated in a recursive fashion based on small SoC segments in \cite{Bakas2017}, but updating of impedance parameters is not considered. On the other hand, in \cite{Jiang2019}, an adaptive Extended Kalman Filter (EKF) is applied to estimate the states, i.e., SoC and overpotential, in combination with two separate LS estimators for estimating model parameters and capacity. Lastly, in \cite{janssen2022}, LS is not applied, but instead, capacity is integrated into the EKF by adding aging dynamics to the model structure. While potentially powerful, it requires significant a priori parametrization effort. Overall, these implementations focus on estimating either capacity or impedance but not both or require significant modeling effort prior to deployment.

In this paper, a Joint Extended Kalman Filter (JEKF) with forgetting factor, estimating SoC, overpotential, and only two model parameters, is paired with an ordinary LS capacity estimator with forgetting factor which is evaluated only once per (dis)charging segment, automatically triggered by cycle detection logic. This coupled structure, constituted of pre-existing structures for separate estimation of impedance and capacity altered with seemingly minor yet important changes, provides computationally inexpensive impedance and capacity estimation while requiring minimal parameterization prior to deployment. The coupled estimation structure is tested on dynamic excitation data measured at several stages of aging for the same cell, and the results demonstrate a consistent relation with impedance increase, capacity decrease, and cell age. Lastly, this structure has been deployed in a Hardware-in-the-Loop (HiL) setting and the results verify the offline performance.

%overview of contents
%This paper is structured as follows. Section \ref{sec:ParameterEstimator} details the method are explained. In Section \ref{sec:proofofconcept} the algorithm is demonstrated on data of differently aged cells. Additionally, in Section \ref{sec:paramestres} the algorithm is implemented on a HiL setup and shown to function on realistic vehicle data. Lastly, conclusions and future work are presented in Section \ref{sec:conclusions}.

\section{Simultaneous State, Parameter, and Capacity Estimation}\label{sec:ParameterEstimator}
Before diving deeper into the capacity and parameter estimation approach, let us first discuss the model structure at hand. In this paper, we consider the battery to be modeled by a first-order dynamical system given by
\begin{subequations}\label{eq:model}
\begin{align} \label{eq:model_a}
    \begin{bmatrix}s_{k+1}\\o_{k+1}\end{bmatrix} &= \begin{bmatrix}1 & 0\\ 0 & \theta_{1}f_1(\pmb{p}_k)\end{bmatrix}\begin{bmatrix}s_{k}\\o_{k}\end{bmatrix} + \begin{bmatrix} \tfrac{\tau}{C_0}\\ \theta_{2}f_2(\pmb{p}_k)\end{bmatrix}u_k,\\
    \hat{y}_{k}&= g(s_k) + o_k+\theta_{3}f_3(\pmb{p}_k)u_k,
\end{align}
\end{subequations}
with $k\in\mathbb{Z}^+$ the time, $s_k$ the SoC, $o_k$ the dynamic part of the overpotential, $u_k$ the applied current and $\hat{y}_{k}$ the predicted terminal voltage. Furthermore, $g(s_{k})$ is a monotonic function representing the Electromotive-Force (EMF), also known as open-circuit voltage, $C_0$ [As] is the battery capacity, $\theta_{1}f_1(\pmb{p}_k)$ is the overpotential relaxation rate, $\theta_{2}f_2(\pmb{p}_k)$ the overpotential increase due to the applied current and $\theta_{3}f_3(\pmb{p}_k)$ models the Ohmic resistance and all high-frequency impedances, i.e., faster than the sampling time $\tau$. Here, $f_i$ with $i\in\{1,2,3\}$ are functions of $\pmb{p}_k=\{p_0,\dots,p_k\}$ representing one or multiple scheduling variables. Essentially, $\theta_i f_i(\pmb{p}_k)$ is a pre-parametrised function multiplied with a scaling factor $\theta_i$. In this way, the dependency of system dynamics on, e.g., SoC or temperature, can be incorporated while maintaining adaptability with respect to aging or model uncertainty. In the remainder of this paper, we assume free system dynamics, i.e., $f_i(\pmb{p}_k)=1$ for all $i$.

Note that system \eqref{eq:model} is fully equivalent to a 1-RC pair equivalent-circuit model, with series resistance $R_0$, and RC pair resistance $R_1$ and capacitance $C_1$, with exact equivalence given by
\vspace{-15pt}
\begin{align}
R_0  &= \;\;\;\theta_{3}, \\
R_1 &= - \theta_{2} / (\theta_{1} - 1), \\
C_1 &= -(e^{-t_s}(\theta_{1} - 1)) / (\theta_{1}\theta_{2}).
\end{align}
Considering system \eqref{eq:model}, we can now formalize the goal of capacity and impedance tracking as
\begin{equation}
   \label{eq:goal} \underset{C_0, \theta_{1}, \theta_{2}, \theta_{3}}{\text{min}} \quad \sum_{k\in\mathcal{K}}(y_k-\hat{y}_{k})^2,
\end{equation}
with $y_k$ the measured battery terminal voltage and samples $\mathcal{K}=\{1,\dots,K\}$ where $K$ marks battery End-of-Life.

Solving \eqref{eq:goal} is non-trivial, let alone doing so in an online recursive fashion, which would be required on the BMS. This paper proposes an estimation structure consisting of two main elements, namely a JEKF with forgetting factor, as presented in \cite{Beelen2021}, which estimates the SoC and the dynamical model parameters and the Recursive Least Squares (RLS) capacity estimator.

\subsection{JEKF State and Parameter Estimation}\label{sec:JEKF}
The JEKF employed here is taken from \cite{Beelen2021}, where the JEKF is combined with a forgetting factor which enables single-knob tuning, i.e., choosing the forgetting factor $\gamma$. The `joint' aspect of the Kalman filter implies that in addition to the internal model states, such as SoC, also model parameters are being estimated. In this case, only $\theta_{2}$ and $\theta_{3}$ will be estimated by the filter, unlike in \cite{Beelen2021}, where all overpotential parameters are estimated. Incorrect estimation of $\theta_1$ can occur in situations with poor observability \cite{Beelen2021}, such as constant-current charging. Due to the severe consequences of incorrect estimation of $\theta_1$ on model stability, namely if $\theta_1>1$ then \eqref{eq:model} is unstable, $\theta_1$ is chosen to be fixed. In this paper, a value of $\theta_1 = 0.99$ is chosen for a model sampling time of $\tau=1$\,s.

To estimate $\theta_{2}$ and $\theta_{3}$ online, system \eqref{eq:model_a} is extended according to
\vspace{-10pt}
\begin{subequations}\label{eq:JEKF_ECM}
\begin{align}
x_{k+1} &= \left[\begin{array}{cc|c}
1 & 0 & 0 \\
0 & \theta_{1} & 0 \\ \hline
0 & 0 & I
\end{array}\right]x_k+\left[\begin{array}{c}
\frac{\tau}{C_0} \\
\theta_{2,k} \\ \hline
0
\end{array}\right]u_k,\\
\hat{y}_{k}&= g(s_k) + o_k+\theta_{3,k}u_k,
\end{align}
\end{subequations}
with $x_k=[s_k,o_k, \theta_{2,k}, \theta_{3,k}]^\top$ and $I=\text{diag}([1,1])$ the identity matrix. In \eqref{eq:JEKF_ECM}, the state vector has been extended with the, now time-varying, model parameters  $\theta_{k+1}=\theta_k$. The dynamics for $\theta_1$ are omitted for reasons mentioned above. The benefit of using \eqref{eq:JEKF_ECM}, is that only limited prior knowledge, in the form of the EMF $g(s_k)$ and an estimate for $\theta_1$ are required as input, while allowing for impedance increase, in the form of $\theta_{2,k}$ and $\theta_{3,k}$, as the battery ages. For details on the JEKF and how to select the forgetting factor $\gamma$, the reader is referred to \cite{Beelen2021}.

\subsection{Recursive Least Squares Capacity Estimation}\label{sec:RLS}
Besides accommodating for impedance increase, it is crucial to track capacity fade by updating $C_0$. In essence, similar to many others such as \cite{Plett2011,Bakas2017}, the capacity is estimated by comparing the difference in estimated SoC and the recorded difference in capacity throughput over a certain window, yielding a capacity estimate according to
\begin{equation}\label{eq:RLS}
\hat{C}_0 = \frac{\tau\sum_{k=a}^{b}u_k}{\hat{s}_{b}-\hat{s}_{a}},
\end{equation}
with $a$ and $b$ denoting the start and end time-instant of the window, respectively. In this paper, we consider a simple fading-memory ordinary least-squares optimization problem given by
\begin{equation}\label{eq:RLS_cost}
    \underset{\hat{C}_0}{\text{min}} \quad \sum^n_{j=1}\lambda^{n-j}\Big(\tau\sum_{k=a_j}^{b_j}u_k - \hat{C}_0(\hat{s}_{b_j}-\hat{s}_{a_j}) \Big)^2,
\end{equation}
where $\lambda\in[0,1]$ is the forgetting factor, $n\in\mathbb{Z}^+$ the number of windows, and $a_j$ and $b_j$ the start and end time-instant of each window $j$. The optimal solution and its corresponding recursive form can be found in, e.g., \cite{plett2015battery}. Due to its simplicity, the ordinary least squares implementation is preferred here over alternatives such as total least squares. The merits of the latter, attributed to more complicated implementations, are often based on assumptions like Gaussian distributed errors in current sensing or SoC estimation. However, in practice these errors are often encountered to be of a more complicated nature such as biased current sensors, such as treated in \cite{Michel2023}, or skewed SoC estimates due to limitations of the underlying model structure \eqref{eq:model} or errors in the applied EMF realization $g(s_k)$. Therefore, and because it is most practical for implementation on embedded hardware, the authors opt for RLS with forgetting factor.

In this paper the window size is maximized by choosing $a$ and $b$ respectively as the start and end of charging sessions, thus reducing the number of evaluations of \eqref{eq:RLS} to one per charging segment. Only continuous charging segments where $(\hat{s}_{b}-\hat{s}_{a}) > 0.2$ are considered. Not only does this result in a computational reduction, but it potentially minimizes the impact of SoC estimation errors on the estimate $\hat{C}_0$, as will be shown in Section~\ref{sec:proofofconcept}. Namely, assuming the integration error of the current sensor is relatively small, the larger the difference between $\hat{s}_a$ and $\hat{s}_b$, the smaller the impact of estimation errors will be. Note that more frequent evaluation is not likely to reduce impact of SoC estimation errors due to its non-zero mean nature.

\section{Results - Single Cell with Aging}\label{sec:proofofconcept}
To demonstrate the performance of the JEKF and the effect of the iterative capacity updates, the proposed algorithm is evaluated over cycling data from an LG M50 21700 cell. For this cell, no other prior information is available except the EMF-curve $g(s_k)$, obtained from a C/20 discharge and its corresponding capacity $C_0$. The data represents 10 WLTP-discharge-CC-charge cycles, which combined last 36 hours. This data is available at a sampling rate of 10Hz, yet is downsampled to 1\,Hz before analysis. Lastly, the same cycling data is available at three stages of the cell's life while it was subject to aging tests: at Beginning of Life (Dec '21), at mid-life (Mar '22) and towards end-of-life (Jun '22). Between these moments, the cell was continuously cycled performing the described drivecycles at an ambient temperature of 25\,\textsuperscript{o}C.

\subsection{JEKF Performance}
To test the performance of the JEKF, as described in Section\,\ref{sec:JEKF}, it is evaluated over the aforementioned beginning-of-life data. The JEKF is initialized with the initial state estimate $\hat{x}_{0} = [s_0,\, 0,\, 10^{-4},\, 0.02]^\top$, where $s_0$ is determined by mapping the first voltage measurement to the SoC via the EMF $g(s_k)$. As implied by the initial overpotential $o_0 = 0$, the experiment is assumed to start in rest. Secondly, the covariance matrix of the JEKF is initialized as $P_0 =  \text{diag}([1,  1, 10^{-6},  4\cdot10^{-4}])$, which indicates the relatively large uncertainty which is placed on this relaxed-start assumption. The single-knob tuning factor is chosen as $\gamma = 0.999$ to allow for relatively fast changes in the parameters $\theta_2$ and $\theta_3$.

\begin{figure}[t]
  \centering
  \includegraphics[width=\columnwidth]{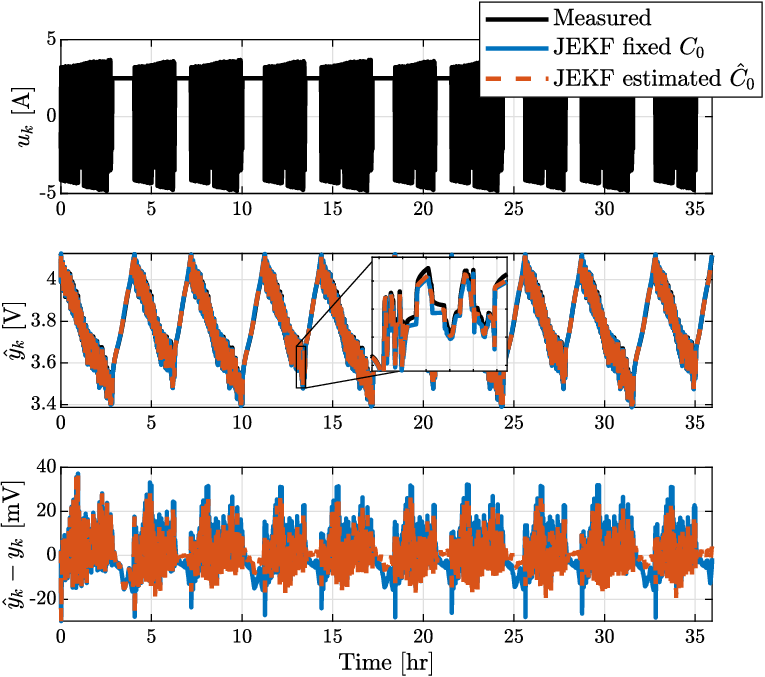}
  \caption{The input current $u_k$ and the estimated terminal voltage $\hat{y}_k$ of the JEKF on 36 hours of WLTP data (beginning-of-life), both \underline{without} and \underline{with} capacity updates.}
  \label{fig:JEKF_IV}
\end{figure}

\begin{figure}[t]
  \centering
  \includegraphics[width=\columnwidth]{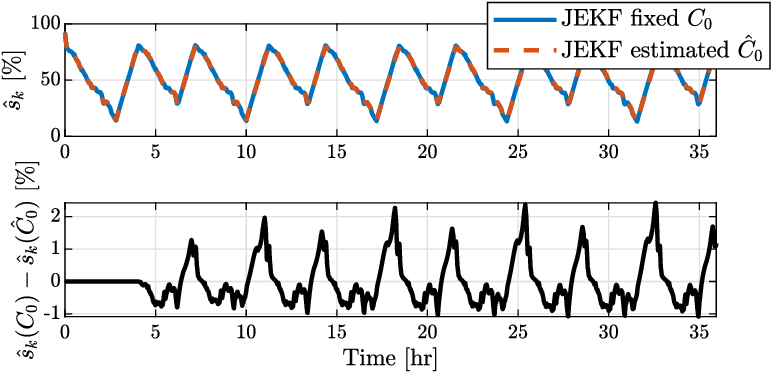}
  \caption{The estimated SOC $\hat{s}_k$ of the JEKF on 36 hours of WLTP data (beginning-of-life), both \underline{without} and \underline{with} capacity updates.}
  \label{fig:JEKF_SOC}
\end{figure}

Evaluating the JEKF over the beginning-of-life data without updating the capacity results in the blue lines in Fig.~\ref{fig:JEKF_IV} and Fig.~\ref{fig:JEKF_SOC}. Here, the results show that the JEKF converges to a plausible SoC value within the first minutes and that the JEKF tracks the terminal voltage of the cell well. The RMS value of the terminal voltage error, visualized in the bottom graph of Fig.~\ref{fig:JEKF_IV}, is 7.1\,mV over the entire dataset.

%\begin{figure}[t]
%  \centering
%  \includegraphics[width=\columnwidth]{./Images/JEKF_WLTP_Zoom.eps}
%  \caption{Zoomed-in version of Fig.~\ref{fig:JEKF}.}
%  \label{fig:JEKF_Zoom}
%\end{figure}

\subsubsection{Capacity Estimation}
Next, the RLS algorithm is used to estimate capacity based on the charging segments, as described in Section~\ref{sec:RLS}. For this, $\lambda = 0.7$ is chosen to enable fast convergence of the RLS algorithm. The resulting capacity estimates are shown in Fig.~\ref{fig:thetas}, subfigure 1, which shows that the estimated capacity starts at the reference value of 4.4\,Ah, and changes after the first CC charge is complete at around $t = 4\,$hr. After this, small changes are applied every time a new CC charge is complete resulting in a final capacity estimate of 4.7\,Ah. The resulting JEKF performance under the influence of these incremental capacity updates is displayed in Fig.~\ref{fig:JEKF_IV} and Fig.~\ref{fig:JEKF_SOC} by the red, dashed line. The results show that initially, the output of both methods is the same, until after the first CC-charge, when the first capacity update occurs. After this moment, a difference in estimated SoC, shown in Fig.~\ref{fig:JEKF_SOC}, starts to appear and the performance of the JEKF with estimated capacity improves, reducing the RMS voltage error to 5.2\,mV.

\subsubsection{Impedance Estimation}
\begin{figure}[!tb]
  \centering
  \includegraphics[width=\columnwidth,trim={0.65cm 0.9cm 1cm 0.85cm},clip]{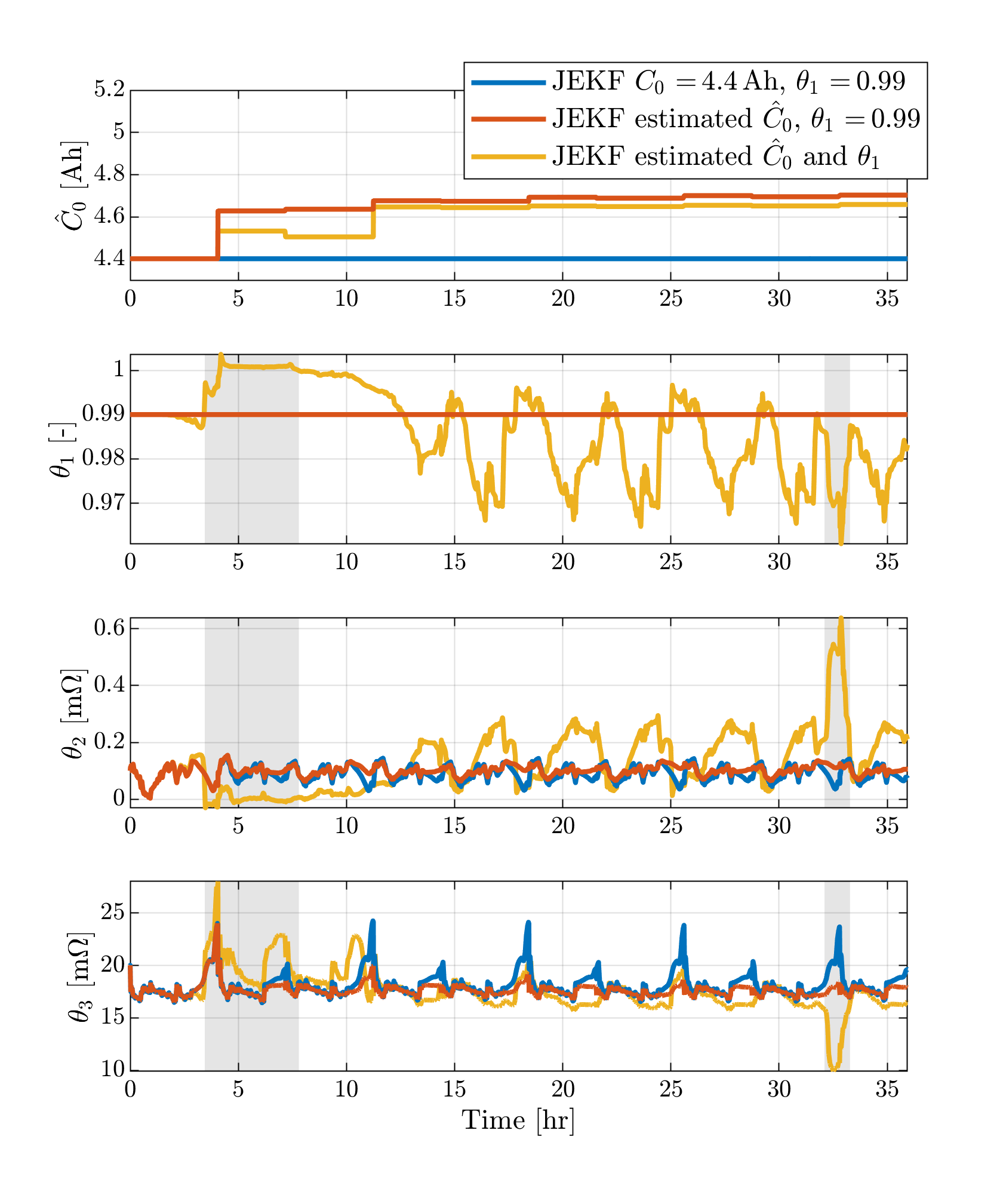}
  \caption{The estimated capacity $\hat{C}_0$ and estimated JEKF parameters $\theta_1$, $\theta_2$, and $\theta_3$ on 36 hours of WLTP data (beginning-of-life), both \underline{without} and \underline{with} capacity updates, and once with variable $\theta_1$. Regions where the JEKF is marginally stable are shaded.}
  \label{fig:thetas}
\end{figure}

Fig.~\ref{fig:thetas} shows the parameter estimates of the JEKF both with and without the incremental capacity updates. In the fixed-capacity case, indicated by the blue line, the impedances $\theta_2$ and $\theta_3$ show evident variation. Most notable are the peaks in $\theta_3$ during the CC charging sections. Without capacity updates, the average value of $\theta_3$ is 18\,m$\Omega$. The parameter estimates for the algorithm with capacity updates are visualized in Fig.~\ref{fig:thetas}, red line, and indicate that $\theta_2$ and $\theta_3$ show less variation, with the the notable peaks in $\theta_3$ disappearing after the first charging segment, when the capacity is first updated. The mean value for $\theta_3$ is 17.6\,m$\Omega$.

Additionally, Fig.~\ref{fig:thetas} shows in yellow the results where $\theta_1$ is estimated as part of the joint state vector, just as originally presented in \cite{Beelen2021}, while also using the RLS algorithm to estimate $\hat{C}_0$. The respective initial covariance value used is $10^{-7}$. The results show that $\theta_1 > 1$ for a moment at $t = 4$\,hr, after completion of the first CC charge. This goes paired with a negative value of $\theta_2$, both of which indicate marginal stability of the JEKF. After this, the values seem to stabilize. However, the deviations of $\theta_2$ and $\theta_3$ at $t = [32, 33.5]$ indicate that impedance parameters are challenging to estimate uniquely, especially during charging. These incorrect parameter values will affect the SoC estimated by the JEKF, and subsequently the capacity values estimated from it. This motivates the initial choice to keep a fixed value for $\theta_1 = 0.99$, which is kept for the remainder of this paper.

Note that the impedance-estimation performance can be further improved by pre-parameterizing functions $f_i(\pmb{p}_k)$ in \eqref{eq:model} to capture the dependency of impedance on the SoC, or additionally but less relevant for this data, on temperature. However, this does require additional a priori testing and parametrization effort.

\subsection{Comparing Differently Aged Cells}
\begin{figure}[t]
  \centering
  \includegraphics[width=\columnwidth]{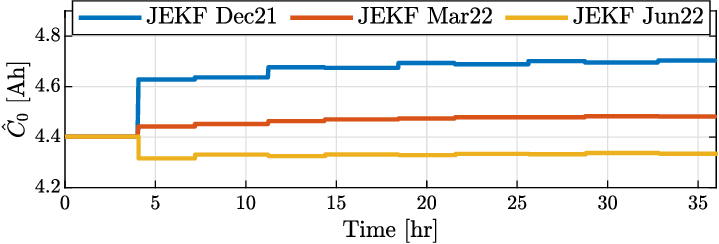}
  \caption{The estimated capacity $\hat{C}_0$ based on 36 hours of WLTP data for three different aging states.}
  \label{fig:JEKF_WLTP_CapRi_Aging}
\end{figure}

In order to test if the battery parameter estimation algorithm adapts to aging behaviour, the three different datasets recorded at the three different aging stages of the cell are evaluated. The hypothesis is that when going from beginning-of-life towards end-of-life, the cell will display a decreasing capacity, represented by $\hat{C}_0$ and increasing resistance, represented by $\theta_2$ and $\theta_3$. The resulting capacity estimates are displayed in Fig.~\ref{fig:JEKF_WLTP_CapRi_Aging} and show a clear distinction between the three different data sources. In all three cases, the first capacity update, after the first CC-charging session at $t = 4\,$hr, establishes an estimate that is close to the final estimates, which are respectively 4.72\,Ah (100\%), 4.48\,Ah (95.0\%), and 4.33\,Ah (91.7\%).
\begin{figure}[!t]
  \centering
  \includegraphics[width=\columnwidth]{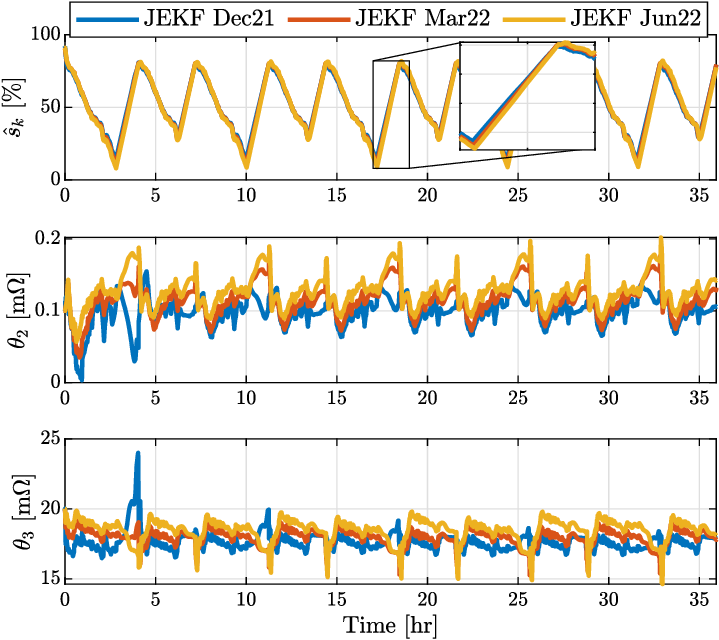}
  \caption{The estimated SoC and JEKF parameters for three different aging states.}
  \label{fig:JEKF_WLTP_RC_Aging}
\end{figure}
Furthermore, differences in the parameter estimates are observed between the different aging states in Fig.~\ref{fig:JEKF_WLTP_RC_Aging}. Firstly, the SoC increase during charging is slightly larger on the aged cell, indicating a decreased capacity. Secondly, the average parameter values, are increasing with battery age. The average values are summarized in Table~\ref{tab:ParameterEstimatesAging}. These confirm a plausible decreasing trend in the observed capacity and an increasing trend in the impedances.

\begin{table}[t]
\centering
\caption{Estimated parameters based on data from different stadia of the cell's life, based on JEKF+RLS. Mentioned capacity is last value, impedance is averaged over 36\,hr.}\label{tab:ParameterEstimatesAging}
\begin{tabular}{|c||c|c|c|}
  \hline
  \bfseries Aging State & \bfseries Capacity $\hat{C}_0$ [Ah] & \bfseries $\theta_2$ [m$\Omega$] & \bfseries $\theta_3$ [m$\Omega$]\\
  \hline
  Beginning-of-life   & 4.72 & 0.10 & 17.6 \\
  Mid-life            & 4.48 & 0.11 & 17.9 \\
  Towards end-of-life & 4.33 & 0.13 & 18.5 \\
  \hline
\end{tabular}
\end{table}

\subsection{Importance of Proper Segment-Selection}
\begin{figure}[t!]
  \centering
  \includegraphics[width=\columnwidth]{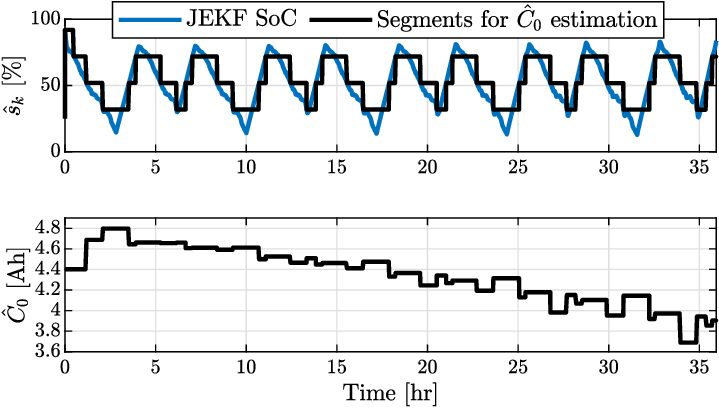}
  \caption{Results of an alternative method where the segments used for capacity estimation are strictly segments of 20\% SoC window. Data is from beginning-of-life.}
  \label{fig:Segments_reference}
\end{figure}

The proper selection of segments of data used to update the capacity estimate is imperative. All of the above results assume the complete charging segments as input to the RLS estimator, which is computationally practical and plausible assumption for an automotive use case. To emphasize the importance of this choice, the same framework, i.e., JEKF with RLS capacity estimation, is demonstrated in Fig.~\ref{fig:Segments_reference}, yet with a different segment choice. In this case, a segment is ended as soon as
\begin{equation}
    |\hat{s}_{b_i}-\hat{s}_{a_i}| > 0.2\;,
\end{equation}
thereby also including discharging segments.
The results indicate a capacity estimate which is more frequently updated, yet diverges from the expected value of 4.72\,Ah.

\section{Hardware-in-the-Loop Implementation}\label{sec:paramestres} %Camiel
\begin{figure}[ht!]
    \centering
    \subfloat[\label{1a}]{%
        \includegraphics[width=0.40\columnwidth]{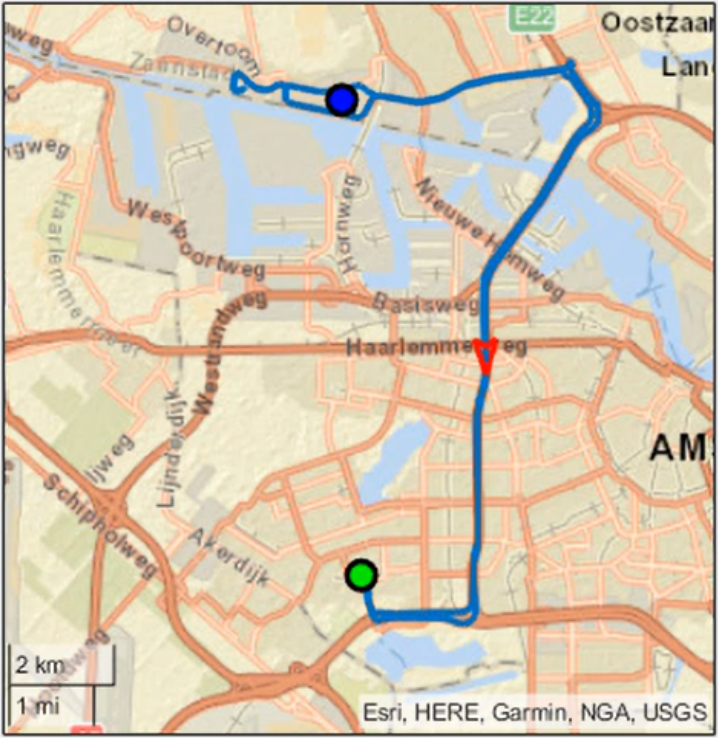}
        \label{fig:first_sub}
    }
    \subfloat[\label{1b}]{%
        \includegraphics[width=0.40\columnwidth]{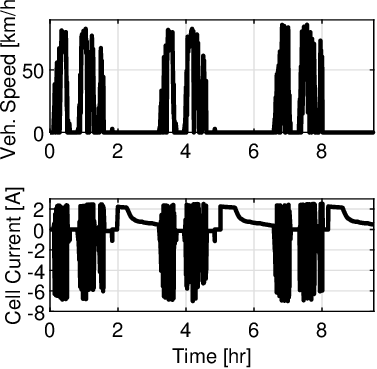}
        \label{fig:second_sub}
    }
    \caption{The simulated route (a) and the resulting velocity profile and cell current (b).}
    \label{fig:HiL_route}
\end{figure}
\begin{figure}[ht!]
\centering
\includegraphics[width=\columnwidth,trim={0.65cm 0.0cm 0.75cm 0.45cm},clip]{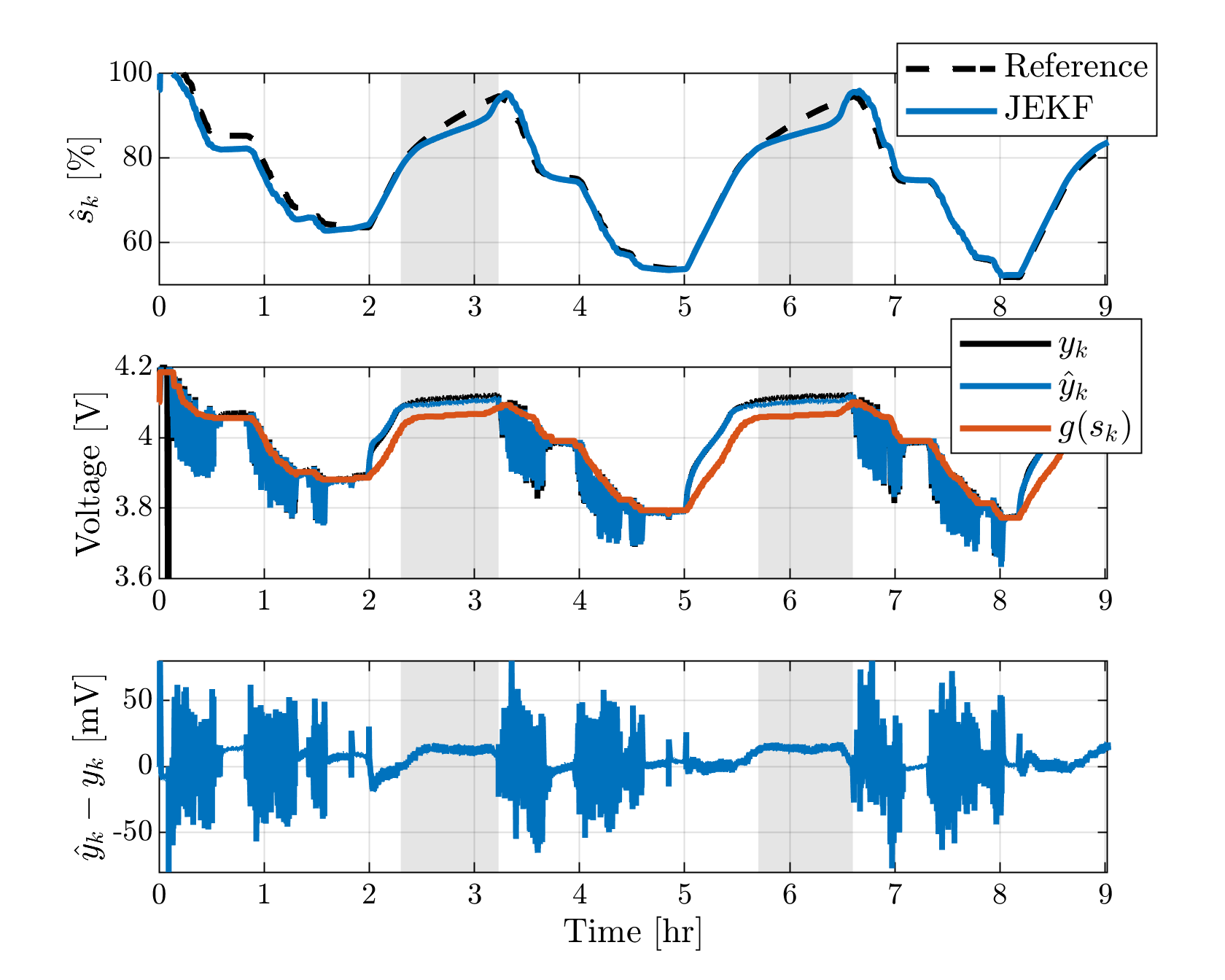}
\caption{The results of the HiL test with the estimated SoC $s_k$, a reference SoC obtained by coulomb-counting, the estimated open-circuit voltage $g(s_k)$, and the estimated terminal voltage $\hat{y}_k$ with respect to the measured terminal voltage $y_k$. The shaded regions indicate the cv-stage of charging.}\label{fig:BPE_HiL_SOC}
\end{figure}
\begin{figure}[t]
  \centering
  \includegraphics[width=\columnwidth,trim={0.65cm 0.0cm 0.75cm 0.45cm},clip]{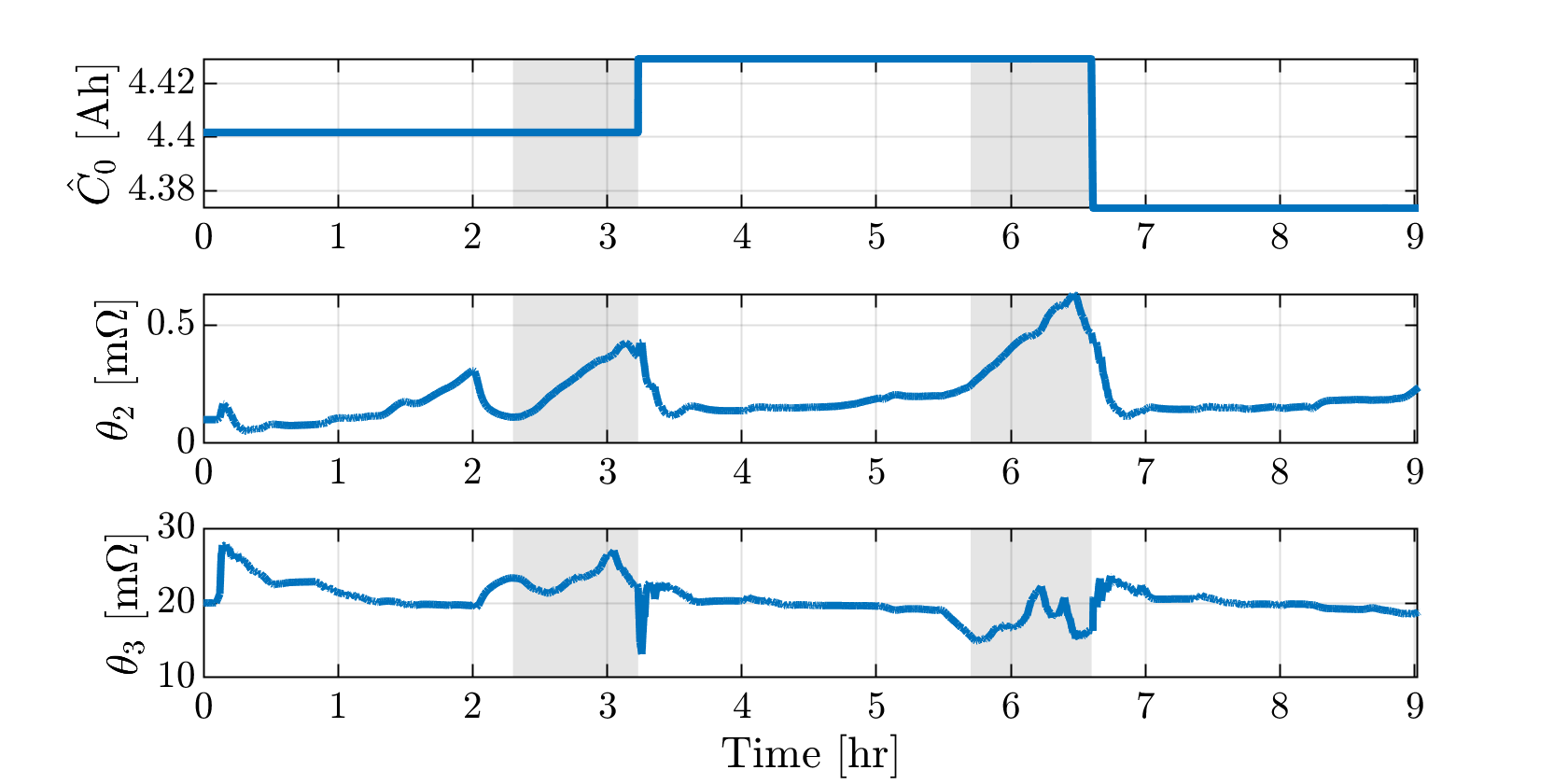}
  \caption{The resulting capacity and impedance parameter estimates during the HiL test. The shaded regions indicate the cv-stage of charging.}
  \label{fig:A1RCvalues}
\end{figure}
To demonstrate the application of the algorithm in real-world operating conditions, the JEKF+RLS algorithm is implemented as an algorithm in the Simulink-based TNO ADVANCE vehicle modeling environment \cite{Tillaart2001}.
The inputs to this control algorithm are current $u_k$ and voltage $y_k$ and the main outputs are JEKF-filtered voltage $\hat{y}_k$, SoC $s_k$, capacity $\hat{C}_0$, $\theta_2$, and $\theta_3$.
The TNO ADVANCE implementation allows the algorithm to be integrated into a vehicle powertrain simulation representing a battery-electric truck.
A Hardware-in-the-Loop (HiL) test is performed, where a cell of the same type as described in Section~\ref{sec:proofofconcept}, at beginning-of-life-conditions, is subjected to drive cycles determined by the vehicle model, and the state estimates performed by the JEKF are used for vehicle control and charging-strategy decisions.
A map of the vehicle route and the resulting current profile are shown in~Fig.~\ref{fig:HiL_route}.

During the HiL test, the purpose of the algorithm is to provide an up-to-date SoC estimate for real-time control as well as to provide a continuously-adapted capacity $\hat{C}_0$ as input to charging strategies. Fig.~\ref{fig:BPE_HiL_SOC} shows the estimated SoC and the corresponding measured and predicted terminal cell voltage. The figure shows that the SoC is generally accurate yet during charging deviations in SoC and terminal voltage can occur during the constant-voltage (cv) stage of charging. During these same periods, the parameter estimates $\theta_2$ in Fig.~\ref{fig:A1RCvalues} are shown to deviate. The drifting of the JEKF is believed to be partly caused by the EMF being relatively flat around the 80\,\% SoC for this particular cell, in combination with the lack of dynamics in the current during charging. As a result, the JEKF estimates increasing impedance parameters instead of increasing the SoC to match the observed terminal voltage. This deviation is corrected as soon as discharging ends. Nevertheless, these deviations during charging will influence the estimated capacity and are considered a topic of future research.

\section{Conclusions}\label{sec:conclusions}
\balance
This paper presents a combined method to estimate both battery capacity and impedance.
The algorithm consists of a JEKF to estimate SoC and impedance, and an RLS algorithm that determines the capacity explicitly based on data from charging segments.
By analyzing the algorithm on cell data with WLTP cycles it is shown that by adapting the capacity the RMS voltage error of the Kalman filter is reduced from 7.1\,mV to 5.2\,mV.
The estimated capacity converges already after one CC charging session and the estimated capacity values follow a decreasing trend as the cell ages; from 4.72\,Ah at beginning-of-life to 4.33\,Ah near end-of-life.
Accordingly, the instantaneous internal resistance of the cell is estimated to increase from 17.6\,m$\Omega$ to 18.5\,m$\Omega$ due to aging.
%
%Proper selection of the segments to base the capacity estimate on is essential, as improper selection is shown to result in diverging estimates.
%
Lastly, the algorithm is demonstrated in a HiL setup with a simulated vehicle and a physical cell.

Future work includes the implementation of the algorithm on embedded hardware and further research on the optimal selection of the capacity-estimation segments.
Additionally, extensions of the algorithm to include temperature-related effects should be considered, for instance by identifying temperature-dependent functions $f_i(\pmb{p}_k)$ in the model description.

%%Bibliography
\bibliographystyle{./IEEEtranBST2/IEEEtran}
%\IEEEtriggeratref{1}
\bibliography{./IEEEtranBST2/IEEEabrv,./Publications-VPPC24_corrected_IEEE}

\begin{thebibliography}{10}
\providecommand{\url}[1]{#1}
\csname url@samestyle\endcsname
\providecommand{\newblock}{\relax}
\providecommand{\bibinfo}[2]{#2}
\providecommand{\BIBentrySTDinterwordspacing}{\spaceskip=0pt\relax}
\providecommand{\BIBentryALTinterwordstretchfactor}{4}
\providecommand{\BIBentryALTinterwordspacing}{\spaceskip=\fontdimen2\font plus
\BIBentryALTinterwordstretchfactor\fontdimen3\font minus
  \fontdimen4\font\relax}
\providecommand{\BIBforeignlanguage}[2]{{%
\expandafter\ifx\csname l@#1\endcsname\relax
\typeout{** WARNING: IEEEtran.bst: No hyphenation pattern has been}%
\typeout{** loaded for the language `#1'. Using the pattern for}%
\typeout{** the default language instead.}%
\else
\language=\csname l@#1\endcsname
\fi
#2}}
\providecommand{\BIBdecl}{\relax}
\BIBdecl

\bibitem{su2022fast}
X.~Su, B.~Sun, J.~Wang, W.~Zhang, S.~Ma, X.~He, and H.~Ruan, ``Fast capacity
  estimation for lithium-ion battery based on online identification of
  low-frequency electrochemical impedance spectroscopy and gaussian process
  regression,'' \emph{Applied Energy}, vol. 322, p. 119516, 2022.

\bibitem{lyu2021partial}
Z.~Lyu, R.~Gao, and X.~Li, ``A partial charging curve-based data-fusion-model
  method for capacity estimation of li-ion battery,'' \emph{Journal of Power
  Sources}, vol. 483, p. 229131, 2021.

\bibitem{Sui2021}
\BIBentryALTinterwordspacing
X.~Sui, S.~He, S.~B. Vilsen, J.~Meng, R.~Teodorescu, and D.-I. Stroe, ``{A
  review of non-probabilistic machine learning-based state of health estimation
  techniques for Lithium-ion battery},'' \emph{Appl. Energy}, vol. 300, p.
  117346, Oct. 2021. [Online]. Available:
  \url{https://doi.org/10.1016/j.apenergy.2021.117346}
\BIBentrySTDinterwordspacing

\bibitem{Wu2023}
\BIBentryALTinterwordspacing
M.~Wu, L.~Wang, and J.~Wu, ``{State of health estimation of the LiFePO4 power
  battery based on the forgetting factor recursive Total Least Squares and the
  temperature correction},'' \emph{Energy}, vol. 282, p. 128437, Nov. 2023.
  [Online]. Available: \url{https://doi.org/10.1016/j.energy.2023.128437}
\BIBentrySTDinterwordspacing

\bibitem{Bakas2017}
E.~Bakas, B.~Rosca, S.~Wilkins, and T.~Donkers, ``{Least-Squares-based Capacity
  Estimation for Lithium-ion Battery Cells},'' in \emph{EEVC 2017}, Geneva,
  Switzerland, Mar. 2017, p.~6.

\bibitem{Jiang2019}
\BIBentryALTinterwordspacing
B.~Jiang, H.~Dai, X.~Wei, and T.~Xu, ``{Joint estimation of lithium-ion battery
  state of charge and capacity within an adaptive variable multi-timescale
  framework considering current measurement offset},'' \emph{Appl. Energy},
  vol. 253, p. 113619, Nov. 2019. [Online]. Available:
  \url{https://doi.org/10.1016/j.apenergy.2019.113619}
\BIBentrySTDinterwordspacing

\bibitem{janssen2022}
\BIBentryALTinterwordspacing
N.~Janssen, ``{Extended State Estimation of a Lithium-Ion Battery Cell Using
  State-of-Health Estimation},'' Eindhoven, 2022, {MSc.} {T}hesis, Eindhoven
  Univ. of Tech. [Online]. Available:
  \url{https://pure.tue.nl/ws/portalfiles/portal/218080877/1008382\_ExtendedStateEstimation.pdf}
\BIBentrySTDinterwordspacing

\bibitem{Beelen2021}
\BIBentryALTinterwordspacing
H.~Beelen, H.~J. Bergveld, and M.~C.~F. Donkers, ``{Joint Estimation of Battery
  Parameters and State of Charge Using an Extended Kalman Filter: A
  Single-Parameter Tuning Approach},'' \emph{IEEE Trans. Control Syst.
  Technol.}, vol.~29, no.~3, pp. 1087--1101, May 2021. [Online]. Available:
  \url{https://doi.org/10.1109/TCST.2020.2992523}
\BIBentrySTDinterwordspacing

\bibitem{Plett2011}
\BIBentryALTinterwordspacing
G.~L. Plett, ``{Recursive approximate weighted total least squares estimation
  of battery cell total capacity},'' \emph{J. Power Sources}, vol. 196, no.~4,
  pp. 2319--2331, Feb. 2011. [Online]. Available:
  \url{https://doi.org/10.1016/j.jpowsour.2010.09.048}
\BIBentrySTDinterwordspacing

\bibitem{plett2015battery}
------, \emph{Battery management systems, Volume II: Equivalent-circuit
  methods}.\hskip 1em plus 0.5em minus 0.4em\relax Artech House, 2015.

\bibitem{Michel2023}
M.~Paul-Henri and H.~Vincent, ``Robust and adaptive online state-of-health and
  state-of-charge estimation of li-ion battery cell,'' in \emph{2023 IEEE
  Vehicle Power and Propulsion Conference (VPPC)}, Milan, Italy, Oct. 2023,
  p.~6.

\bibitem{Tillaart2001}
\BIBentryALTinterwordspacing
E.~van~den Tillaart, S.~Mourad, and H.~Lupker, ``{TNO} {ADVANCE}: a modular
  simulation tool for combined chassis and powertrain analysis,'' in \emph{All
  Electric Combat Vehicle (AECV) Conference}, Noordwijkerhout, The Netherlands,
  Jan. 2002. [Online]. Available:
  \url{https://resolver.tno.nl/uuid:eee2b103-d626-4847-8af9-23eb7c5f0374}
\BIBentrySTDinterwordspacing

\end{thebibliography}

\end{document}